\documentclass[manuscript]{acmart}

\AtBeginDocument{%
  }

\setcopyright{acmcopyright}
\copyrightyear{2024}
\acmYear{2024}
\acmDOI{XXXXXXX.XXXXXXX}

\acmConference[Conference acronym 'XX]{Make sure to enter the correct
  conference title from your rights confirmation emai}{June 03--05,
  2024}{Woodstock, NY}
\acmPrice{15.00}
\acmISBN{978-1-4503-XXXX-X/24/06}

\acmSubmissionID{1287}

\usepackage{enumitem}

\usepackage{listings}

\usepackage{xcolor} %
\definecolor{attrcolor}{RGB}{100,200,150}
\definecolor{stringcolor}{RGB}{66, 94, 191}

\lstdefinelanguage{query}{
  keywords={compare, identify, trend, filter, attr, op, value, operation, type, name, choice, span, name, range},
  keywordstyle=\color{attrcolor}\bfseries,
  morekeywords={},
  ndkeywords={},
  ndkeywordstyle=\color{black}\bfseries,
  identifierstyle=\color{black},
  sensitive=false,
  comment=[l]{//},
  morecomment=[s]{/*}{*/},
  commentstyle=\color{purple}\ttfamily,
  stringstyle=\color{stringcolor}\ttfamily,
  morestring=[b]',
  morestring=[b]"
}

\lstdefinelanguage{inlinequery}{
  keywords={compare, identify, trend, filter, attr, op, value, operation, type, name, choice, span, name, range},
  keywordstyle=\color{attrcolor}\bfseries,
  morekeywords={},
  ndkeywords={},
  ndkeywordstyle=\color{black}\bfseries,
  identifierstyle=\color{black},
  sensitive=false,
  comment=[l]{//},
  morecomment=[s]{/*}{*/},
  commentstyle=\color{purple}\ttfamily,
  stringstyle=\color{stringcolor}\ttfamily,
  morestring=[b]',
  morestring=[b]"
}

\usepackage[normalem]{ulem}
\usepackage{color}
\usepackage{soul}
\usepackage{tabularx}
\usepackage{array}

\newcommand{\revision}[1]{\leavevmode{\color{blue}{#1}}}
\newcommand{\remark}[1]{\textcolor[RGB]{150, 200, 0}{#1}}

\setstcolor{red}
\newcommand{\removed}[1]{\leavevmode{\color{red}{\st{#1}}}}

\def \cleanversion{} %
\ifx\cleanversion\undefined
\else
 \renewcommand{\remark}[1]{} %
 \renewcommand{\removed}[1]{} 
 \renewcommand{\revision}[1]{#1}
\fi

\begin{document}

\title{Breathing New Life into Existing Visualizations:\\A Natural Language-Driven Manipulation Framework}

\author{Can Liu}
\email{can.liu@pku.edu.cn}
\orcid{1234-5678-9012}
\author{Jiacheng Yu}
\email{jiachengyu@pku.edu.cn}
\author{Yuhan Guo}
\email{yuhan.guo@pku.edu.cn}
\author{Jiayi Zhuang}
\email{wybxc@stu.pku.edu.cn}
\author{Yuchu Luo}
\email{luoyuchu1999@gmail.com}
\author{Xiaoru Yuan}
\email{xiaoru.yuan@pku.edu.cn}
\affiliation{%
  \institution{Peking University}
  \city{Beijing}
  \country{China}
  \postcode{100871}
}

\renewcommand{\shortauthors}{Liu et al.}

\begin{abstract}
We propose an approach to manipulate existing interactive visualizations to answer users' natural language queries. We analyze the natural language tasks and propose a design space of a hierarchical task structure, which allows for a systematic decomposition of complex queries. We introduce a four-level visualization manipulation space to facilitate in-situ manipulations for visualizations, enabling a fine-grained control over the visualization elements. Our methods comprise two essential components: the natural language-to-task translator and the visualization manipulation parser. The natural language-to-task translator employs advanced NLP techniques to extract structured, hierarchical tasks from natural language queries, even those with varying degrees of ambiguity. The visualization manipulation parser leverages the hierarchical task structure to streamline these tasks into a sequence of atomic visualization manipulations. To illustrate the effectiveness of our approach, we provide real-world examples and experimental results. 
The evaluation highlights the precision of our natural language parsing capabilities and underscores the smooth transformation of visualization manipulations.
\end{abstract}



\maketitle

\section{Introduction}

Data visualization is a powerful tool for data exploration and insight communication. Many visualization authoring tools (e.g., D3~\cite{bostock2011d3} and Vega-Lite~\cite{satyanarayan2017vegalite}) empower users to create interactive visualizations for specific predefined tasks. However, enabling natural and free user interaction with visualizations poses two main challenges for both creators and users.
Firstly, creators may find it difficult to anticipate the diverse and complex tasks that users may want to perform on visualizations and to design appropriate and sufficient interaction functions accordingly. The wide range of potential user needs and behaviors can be challenging to fully cover in the design process.
Secondly, users may lack the knowledge or skills to interact with the visualization effectively. Without proper guidance or intuitive interaction design, users may struggle to fully leverage the interactive capabilities of the visualization.
Consequently, many visualizations available to users are either static or limited in interactivity.

To ease user interaction with visualizations, we introduce the use of natural language as an interface.
A natural language interface enables users to express their tasks directly, which is one of the most intuitive forms of interaction.
Moreover, natural language processing technologies have advanced significantly in recent years (e.g., ChatGPT~\cite{openai_chatgpt_api}), mitigating the challenges in natural language processing and understanding.
Introducing natural language interfaces (NLIs~\cite{Srinivasan2017NLI,nli4vis}) for visualizations has become a popular trend, with methods proposed for generating~\cite{yu2020flowsense, luo2021nl2vis}, interacting~\cite{srinivasan2020inchorus}, and editing~\cite{wang2023authoring} visualizations using NLIs. 
However, current visualization NLI implementations often rely on imperative language~\cite{CollectingCharacterizingNLU}, where users directly specify a command to construct visualization or interact with visualization elements. 
These imperative languages are built according to specific syntactic rules, which necessitate users to learn and memorize these rules and have sufficient experience in using them in order to write commands that accurately express their intentions.
Aiming to facilitate natural and direct manipulation with visualizations for diverse tasks, this method addresses two aspects.
First, our method lowers the difficulty of user interaction by enabling users to express their tasks directly.
Second, it proposes a generalized method that can adapt to visualizations from various topics created by different methods.

We propose a method that enables manipulating existing visualizations to respond to natural language queries, regardless of the visualizations' original design and implementation. We aim to augment existing visualizations by seamlessly integrating interactivity implementing dynamic manipulations and re-encoding strategies to adapt to a wide array of visualization tasks. To tackle the aforementioned challenges, we undertake the collection and organization of potential natural language inquiries related to visualizations. Additionally, we propose a design space for representing visualization-related tasks. Within this framework, we introduce a deep learning-based natural language-to-task translator (NL-task translator) specifically engineered to parse natural language queries into structured and hierarchical task descriptions.
To train the NL-Task translator, we leverage large-scale language models (such as GPT3.5) to assist in curating a diverse cross-domain dataset comprising natural language expressions and their associated tasks. Then, we use this dataset to train a smaller model, balancing affordability and accuracy. This process can be seen as a form of knowledge distillation, where the knowledge of the large model is used to guide the learning of the smaller model.
Once we have successfully extracted hierarchical tasks, we proceed to translate them into concrete visualization manipulations. Furthermore, we define a visualization manipulation space encompassing common visualization types, such as bar charts, line charts, and area charts. This visualization manipulation space encompasses four levels and seven types. The manipulations support dynamically transforming the visual elements of visualizations, aligning them with users' exploration requirements.
By introducing the idea of using a large LLM to assist in dataset construction and training a small LLM, we can reduce the scale and computational overhead of the model while ensuring its performance, making it more cost-effective. This method combines the advantages of knowledge engineering and data-driven approaches, providing a feasible solution for natural language-driven visualization interaction. The contributions of this paper can be summarized as follows:
\begin{itemize}
    \item We proposed a deep learning-based natural language-to-task translator that supports parsing users' natural language queries about visualizations into structured-format tasks.
    \item We curated a dataset for natural language-to-visualization tasks, covering various domains, diverse tasks, and varied natural language expressions.
    \item We proposed a manipulation space for common visualizations and a method for converting visual tasks into a series of visualization manipulations.
\end{itemize}

\begin{figure*}[!ht]
    \centering
    \includegraphics[width=\textwidth]{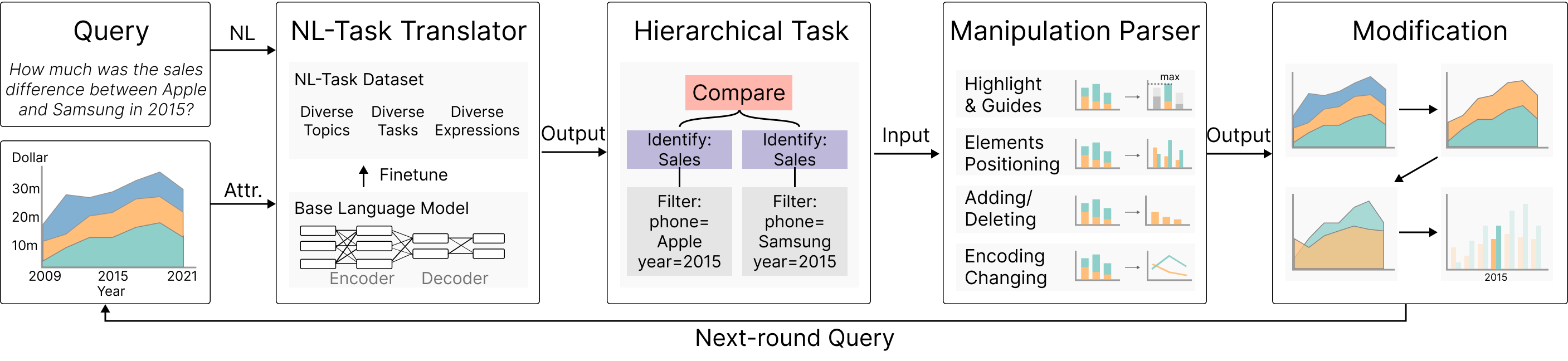}
    \caption{NL-Task translator, fine-tuned on a large language model and a constructed multi-domain and diverse NL-Task dataset, can transform a natural language query input into a hierarchical structure of tasks. The tasks are then transformed into a series of visualization manipulations by the visualization manipulation parser. Finally, the visualization manipulation parser changes the visualization in situ to respond to the natural language query.}
    \label{fig:pipeline}
\end{figure*}

\section{Related Work}
\label{section:background}

Our work provides a natural language interface for performing manipulations on visualizations, which relates to natural language interfaces, visualization tasks, and visualization manipulations.

\subsection{NL Interface for Visualization}

Over the past decades, the field of natural language interfaces (NLIs) for visualization~\cite{nli4vis, deepeye} has attracted significant attention from researchers.
Cox et al.~\cite{cox2001multi} introduced the first pipeline for constructing visualizations from natural language input. To address the inherent ambiguity of natural language, machine learning-based methods~\cite{sun2010articulate, articulate2} and interactive-based methods~\cite{gao2015datatone} have been proposed.
Articulate~\cite{sun2010articulate} employs machine learning techniques to classify visualization tasks based on the words in sentences. DataTone~\cite{gao2015datatone} introduces ambiguity widgets that facilitate interaction to reduce ambiguity.
Shi et al.~\cite{shi2021talk2data} proposed a method for decomposing high-level questions into simpler components, enabling the generation of visual answers.
Previous methods primarily rely on rule-based or deep-learning-based techniques at the word level, but they struggle to handle the large diversity present at the sentence level.
In recent years, large language models (LLM)~\cite{wolf2020transformers} (e.g., BERT~\cite{devlin2018bert}, T5~\cite{raffel2020exploring}, GPT-3~\cite{gpt3}, ChatGPT~\cite{openai_chatgpt_api}) have demonstrated impressive capabilities for adapting to various tasks by fine-tuning on small datasets.
In the field of visualization, several approaches~\cite{liu2021advisor, luo2021natural} focus on parsing natural language using LLMs to construct visualizations.

While the approaches mentioned above focus on constructing visualizations from data, some methods~\cite{setlur2016eviza, hoque2017evizeon, chartQA, Lai2020Annotation, kahou2017figureqa} focused on existing visualizations.
Eviza~\cite{setlur2016eviza} and Evizeon~\cite{hoque2017evizeon} convert natural language input into filters applied to visualizations.
Kim et al.~\cite{chartQA} generate explanations to answer questions related to existing visualization charts.
Some methods aim to generate natural language content for existing visualizations, for example, generate description~\cite{liu2020autocaption} and title~\cite{liu2023autotitle}.
Instead of producing explanations, Lai et al.~\cite{Lai2020Annotation} emphasize highlighting charts to help users better understand them.

Existing approaches mainly concentrate on either QA systems based on visualizations or the construction of visualizations from known data and programming visualization.
In this work, we propose a method that focuses on manipulating existing visualizations that do not depend on underlying data or implementation methods.
Moreover, the natural language queries in this method are not limited to command-based language; instead, they are grounded in diverse, task-oriented questions.

\subsection{Visualization Tasks}

Our goal is to make it possible for users to express their desired visualization tasks in natural language, without needing to specify visualization commands.
We achieve this by mapping natural language input to visualization tasks, which are then translated into visualization manipulations.
To accomplish this, we need to clarify the different levels of the visualization tasks taxonomy.

Brehmer and Munzner~\cite{brehmer2013multi} classify visualization tasks on multiple levels. The "why" level involves users searching for elements of interest (corresponding to data items) and querying on these data items. Queries can include identifying, comparing, and summarizing. Amar et al.~\cite{amar2005lowlevel} provides a more detailed, low-level classification by summarizing ten tasks: retrieving a value, filtering, computing derived values, finding extrema, sorting, determining a range, characterizing distribution, finding anomalies, clustering, and correlating. NL4DV~\cite{narechania2020nl4dv} also classifies tasks into several types. Articulate~\cite{sun2010articulate} classifies natural language words into eight task categories, including comparison, relationship, composition, distribution, statistics, manipulation, and time series. Based on Amar et al.'s taxonomy~\cite{amar2005lowlevel}, Fu et al.~\cite{fu2020quda} construct a natural language utterances tasks classification. Compared to these works, our work focuses on the \textbf{hierarchical structure} of tasks in natural language, such as filtering being a step in value retrieval.

\subsection{Visualization Manipulations}

The manipulation of elements in a visualization aims to alter the visual representations to achieve various user intents.
Yi et al.~\cite{yi2007toward} proposed a multi-level categorization of visualization interactions based on user intent.
Brehmer and Munzner~\cite{brehmer2013multi, munzner2014visualization} classified the manipulation of existing visual elements into six categories, namely, selecting, navigating, arranging, changing, filtering, and aggregating.
Selecting and filtering reduce the focused elements while reconfiguring the spatial layout of visual elements. Changing and aggregating manipulations may involve altering the encoding and abstraction levels. Most of these manipulations can be accomplished through fluent changes of visual elements. 
Manipulation in visualization also reflects in visualization transitions~\cite{heer2007animated} and data videos~\cite{amini2015understanddatavideo, dataplayer}.
Sedig and Parsons~\cite{sedig2013interaction} presented another taxonomy of manipulations on visual elements, which includes unipolar and bipolar actions. Harper and Agrawala~\cite{harper2014deconstructing} deconstructed existing D3 visualizations by matching the given data with visual attributes. Moreover, Harper and Agrawala~\cite{harper2018converting} extracted D3 visualizations and converted them into templates to facilitate reuse.
Lu et al.~\cite{lu2017interaction} extract visual attributes and allow users to interact with existing visualizations by filtering on these visual attributes.
Liu et al.~\cite{liu2023spatialconstraint} proposed a spatial constraint-based model for manipulating static visualizations, which focus on spatial channels (e.g., shape, size, and position).
Our approach aims to leverage visualization manipulations to support users' tasks. We enable not only direct manipulations of visual elements in visualization but also alternative visual representations by adding and removing elements.

\section{Method Overview}
\label{section:overview}

We aim to enable manipulations of static visualization to answer natural language queries.
Command-based natural language queries~\cite{srinivasan2020inchorus, wang2023authoring} have been previously addressed.
Our approach focuses on tasks-based natural language, allowing users to express their intended tasks without specifying the specific changes to visual elements in a visualization.
For instance, a command-based instruction might be ``Sort the countries in the axis according to their heights,'' while a task-based instruction might be ``What is the country with the third highest GDP?''
The latter is more user-friendly as it directly conveys users' thoughts, while the former requires users to have certain visualization expertise.

We first convert natural language into structured visualization tasks. 
Brehmer and Munzner~\cite{brehmer2013multi} classified visualization tasks into three types: identification, comparison, and summarization. 
However, this coarse-grained classification is insufficient, as natural language can convey more intricate structures, such as filters and derivations, which may underlie the hierarchical tasks. 
Users can easily express hierarchical tasks in natural language, but there is a considerable gap between the language and the relevant visualization manipulations due to two factors:
\begin{itemize}[leftmargin=*]
    \item the weak correspondence between natural language and tasks, which is complicated by the ambiguity of natural language and the diversity of users' knowledge backgrounds;
    \item the complex relationship between tasks and visualization manipulations, as the appropriate manipulations for the same task can vary depending on the type of visualization.
\end{itemize}
Conventional methods that depend on templates and rules are inadequate to cope with the diversity of formats present in natural language.
Hence, we adopted a deep learning-based method to extract visualization tasks from natural language input. 

The advancement of large language models (LLMs) has greatly simplified the process of extracting structured information from natural language.
Our goal is to enable lightweight deployment of this process on local computers.
To achieve this, we propose a knowledge distillation approach that combines our domain knowledge with the capabilities of large-size LLMs to curate datasets and fine-tune a smaller LLM. This method allows us to balance accuracy and cost-effectiveness.

Prior to doing so, we delineated the design space of visualization tasks and devised a set of manipulations for common visualizations, such as bar charts, line charts, and area charts, which are among the most prevalent~\cite{battle2018beagle}.
The subsequent sections will elaborate on the design space of visualization tasks (\autoref{section: nl}) and the visualization manipulations (\autoref{section: manipulation}).

\begin{figure*}[!ht]
    \centering
    \includegraphics[width=\textwidth]{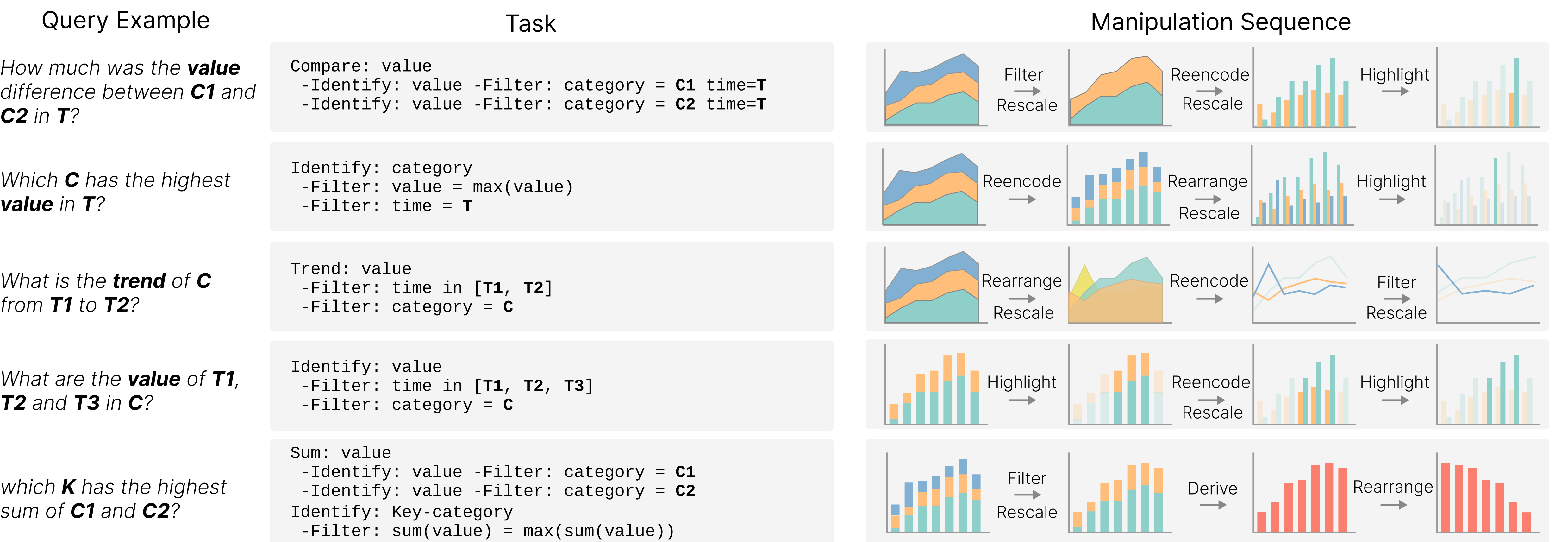}
    \caption{Natural language queries in the context of visualization are transformed into nested high-level tasks, which are represented through a series of visualization manipulations. Visualization manipulations parse tasks from bottom to top, beginning with the resolution of filtering conditions and followed by comparative tasks. Different tasks are represented through the combination of visualization manipulations such as highlighting, annotation, reordering, or remapping.
    }
    \label{fig:querytaskoperation}
\end{figure*}

\section{Tasks Parsing from NL Input}
\label{section: nl}

This section discusses the process of converting natural language into visualization tasks.
Before discussing the method for converting natural language into visualization tasks, we first define the design space of tasks.

We anticipate that the work presented in this paper can be applied to the visualization of data from various domains. We have constructed a cross-domain dataset of natural language-task pairs based on a task-oriented design space and trained our model on it. The dataset construction process involves rule-based predefinition and diversification using a large language model (LLM).

\subsection{Design Space of Visualization Tasks}

Visualization tasks can be classified into three categories: identification, comparison, and summarization~\cite{brehmer2013multi}.
Identification involves finding data items based on known indices or attributes. 
Comparison involves comparing multiple groups of data items.
Summarization involves obtaining overall insights from the visualization.
These three categories reflect the users’ high-level goals, but they omit the low-level manipulations that users execute to accomplish these goals.
High-level tasks such as identification, comparison, and summarization correspond to the users’ primary intentions, but the natural language tasks often exhibit more complexity than this simple classification implies. For instance, identification might entail applying filters or deriving new attributes from existing filters. The comparison might necessitate selecting comparable entities and defining common or distinct filters. Summarization might demand aggregating or sorting data items according to some criteria.

High-level tasks often contain some low-level tasks, such as filtering and derivation.
For instance, the sentence “In 2000, how much was the salary gap between basketball and football?” constitutes a comparison task that incorporates identification.
Filtering enables identifying the visual elements to be considered, while derivation entails calculating new attributes based on the original attributes.
Filtering and derivation can be nested or combined to create a high-level structure.
Filtering can also be applied on top of derivation, for example by identifying data elements based on their rank in a certain attribute.
The above example involves filtering by time (In 2000) and category (basketball and football), as well as deriving new data from their difference.
Hence, a more elaborate and structured approach to describing visualization tasks is required to comprehend how users interact with visualizations and how to devise effective visualization techniques.

\begin{figure}[!ht]
  \centering
  \includegraphics[width=\columnwidth]{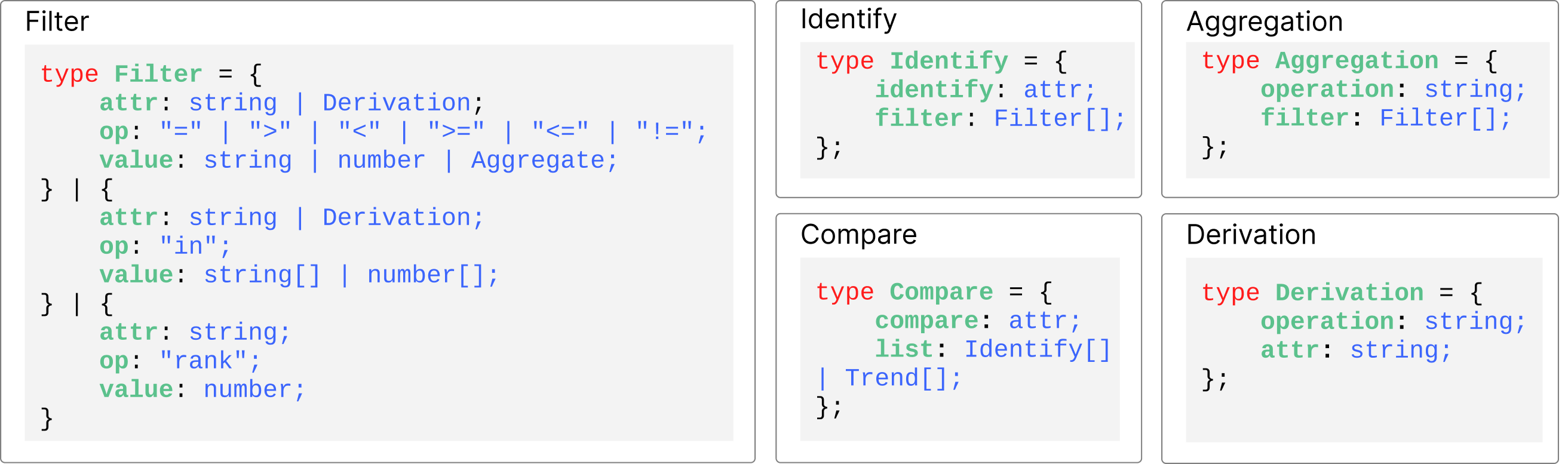}
  \caption{Design space of the natural language task.}
  \label{fig:task_specification}
\end{figure}

Kim et al.~\cite{chartQA} established that more than half of the inquiries related to visualization exhibit a composite structure. Traditional studies~\cite{amar2005lowlevel, narechania2020nl4dv,fu2020quda} typically categorize natural language expressions into paralleled several classes; however, they do not account for the nested structure present in natural language queries. The authors attempt to design structured representations of natural language tasks by employing a set of fundamental operational components. These basic operations within the hierarchical framework encompass filtering, identification, comparison, aggregation, and derivation as shown in \autoref{fig:task_specification}. The \textbf{hierarchical structure} of visualization tasks can thus be constructed utilizing the following fundamental operations.
\begin{itemize}[leftmargin=*]
    \item \textbf{Filtering} is the process of reducing the number of focused data items by restricting the range of visual elements based on certain conditions.
    This can be accomplished by selecting from a categorical list or specifying a range of ordinal or temporal values.
    The general form of filtering is \texttt{\footnotesize{\{attr: "Name", op: "Op", value: "Value"\}}}, for example, \texttt{\footnotesize{\{attr: "time", op: "=", value: "2000"\}}} denotes selecting data items with the attribute time equal to 2000.
    \item \textbf{Identification:} An instance of identification is “What is the price of apples in 2022”.
    The specific identification information pertains to lower-level operations and is recorded in \texttt{\footnotesize{filter}}.
    \item \textbf{Comparison:} An instance of comparison is “What is the difference in price between apples and oranges in 2022?”, which includes two identifications for apples and oranges in 2022 respectively. 
    The sub-attributes of the comparison task are two identification tasks, denoting the objects of comparison.
    \item \textbf{Aggregation:} Aggregation includes max, min, and average to generate a value from a list of data, which can serve as a filter value or an identified value. To determine aggregation, two components are needed: the attribute and the type of aggregation. For instance, to find countries with a life expectancy higher than the average, the syntax of aggregation is \texttt{\footnotesize{\{aggregate: "avg", attribute: "life expectancy"\}}}.
    \item \textbf{Derivation:} The form of derivation generates new attributes according to original attributes, for example, generates rank according to a certain quantitative attribute.
\end{itemize}

This hierarchical structure reflects the nested property of natural language.
For example, the structure of the natural language \textit{``What is the energy type with the fastest percentage growth from 2010 to 2020''} is \texttt{\footnotesize{\{identify: "energy type", filter: [\{ attr: "percentage", op: "=", value: \{aggregate: "max", attribute: "percentage"\}, \{attr: "time", op: "in", value: ["2010", "2020"]\}\}]\}}}.
Such a nested structure necessitates that the method has the ability to precisely parse natural language.
The model needs to identify the content of the result, including the output attribute, and whether comparison and derivation are involved in natural language.

\subsection{Training Data Construction}

The primary goal of the constructed dataset is to train a model capable of handling data attributes across \textbf{diverse domains} and addressing various tasks posed by different users. 
To achieve this, we create a training dataset to ensure diversity in data attributes, tasks, and natural language.
\begin{itemize}[leftmargin=*]
\item \textbf{R1. Task diversity} necessitates that the dataset encompasses operations related to various tasks such as identification, comparison, and summarization.
\item \textbf{R2. Attribute diversity} aims to support visualizations across multiple domains, as the natural language employed to describe attributes from distinct domains may vary significantly.
For instance, the language used to describe fruit prices may differ from that used to discuss daily new cases of specific diseases.
\item \textbf{R3. Visual channel diversity} means allowing users to specify visual elements beyond just data attributes in visualizations, such as color, orientation, and shape.
\item \textbf{R4. Natural language diversity} is crucial for ensuring the model's generalization across different scenarios. As users may have diverse presentation preferences and use various forms of natural language to convey identical meanings, accommodating this diversity in the dataset will improve the model's applicability to a broader range of users and situations.

\end{itemize}

\begin{figure}[!ht]
  \centering
  \includegraphics[width=\columnwidth]{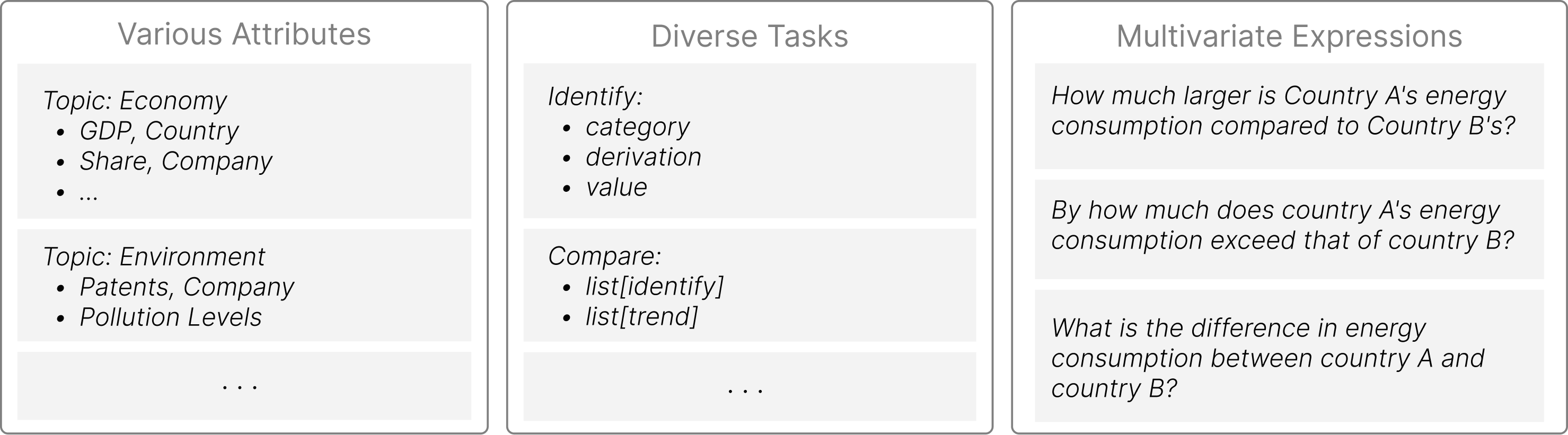}
  \caption{The data creation process ensures diversity in three dimensions, namely data attributes, tasks, and natural language expressions.}
  \label{fig:dataconstruction}
\end{figure}

To meet these requirements, as shown in \autoref{fig:dataconstruction}, we have developed a multi-task dataset using a three-step strategy. The existing NL2SQL datasets~\cite{zhongSeq2SQL2017, xu2017sqlnet, yu2018spider} contain a set of attributes with semantic relationships, such as country and population or football player and score.
The natural language dataset from NL2SQL includes data tables from a variety of domains. In order to ensure the diversity of data attributes and maintain semantic meaning, our dataset also needs to collect data attributes from various domains.
Similar to Spider~\cite{yu2018spider} and WikiSQL~\cite{zhongSeq2SQL2017}, the construction of our dataset consists of three components, namely collecting cross-domain data attributes, generating initial natural language queries using templates, and rephrasing the natural language to make it diverse.

Cross-domain data attributes are combined to encode the content of a visualization.
For instance, a multi-line graph depicting the GDP of multiple countries over several years is a combination of categorical, temporal, and quantitative attributes.
We aim to obtain data attribute combinations from various domains to enable models to analyze data attributes in different scenarios. We collected 486 different data attribute combinations from 65 most frequently discussed topics, such as politics, economics, and environmental protection.
We employ the large language model GPT 3.5 (ChatGPT)~\cite{openai_chatgpt_api} to write down data attributes by providing example data and domains.
The data generated by the large language model resembles real data and is uniformly distributed across different domains.
Below is an example of an attribute combination of data in the field of computer science:
\begin{lstlisting}[language=query]
  {
    attr: [{
      type: "categorical",
      name: "programming language",
      choice: ["Python", "Java", "C++"]
    },{
      type: "temporal",
      span: ["2000", "2021"]
    },{
      type: "quantitative",
      name: "number of users",
      range: ["1000", "1000000"]
    }]
  }
\end{lstlisting}
  
\textbf{Multivariate Tasks}. We have categorized natural language description tasks into several types, including identification, comparison, trend analysis, and summation. Each type of task is further subdivided into specific task categories based on the attributes involved. For the natural language query ``what is the programming language with the most users,'' the task should be as follows.

\begin{lstlisting}[language=query]
  {
    identify: "programming language",
    filter: [{
      attr: "number of users",
      op: "=",
      value: {
        operation: "max",
        attr: "number_of_users"
      }
    }]
  }
\end{lstlisting}

  \textbf{Diversifying Natural Language Expressions.}
  Some of the attribute combinations and initial templates are shown in Table~\ref{tab:nld-tasks}.
  The query format constructed based on templates is often constrained and deviates from real user natural language inputs.
  To tackle this problem, we employ a large language model to rephrase the constructed queries and produce a variety of natural language expressions that align with common language usage.
  For instance, for the following query constructed based on templates: \textit{``Which industry has the highest Revenue in 2015Q1?''},
  the following expressions can be used to represent:
  \begin{itemize}[leftmargin=*]
    \item \textit{What is the industry with the highest earnings in 2015Q1?}
    \item \textit{Which industry generated the most income in the first quarter of 2015?}
    \item \textit{In 2015Q1, which industry had the greatest income?}
    \item \textit{Which sector earned the most revenue in the first quarter of 2015?}
    \item \textit{In 2015Q1, what was the industry with the highest revenue stream?}
    \end{itemize}

We finally constructed a dataset with 5867 pairs of natural language to hierarchical tasks, which are divided into four different sub-tasks: identification (56.60\%), comparison (14.06\%), aggregation (14.11\%), and derivation analysis (15.22\%).
Since attribute names across different domains are diverse while visual channels tend to have a relatively uniform representation, attributes in most pairs (86.65\%) are referred to by their names, with the left 9.32\% referred to by visual channels and 4.02\% by a mix of names and visuals.
On average, there are 2.79 filters per sentence.

\begin{table}[ht]
  \centering
  \caption{Various combinations of attribute types in visualization charts along with their respective natural language query tasks and templates.
  These tasks are broadly categorized into four types, namely identification, comparison, trend analysis, and summation. In the table, the quantitative, categorical, and temporal attributes are referred to as $Q name$, $C name$, and $T name$, respectively. The value choices for these attributes are denoted as $Q_i$, $C_i$, and $T_i$. For instance, if a user wants to determine the maximum stock price of a company at a specific point in time, they may ask a question such as, ``What is the stock price of Apple on Jan 1, 2022?"}
  \label{tab:nld-tasks}
  \begin{tabular}{ccc}
  \hline
  \textbf{Attr.}& \textbf{Category} & \textbf{Question Example} \\
  \hline
  CQ & Identification & What is the \{${Q\text{ name}}$\} of \{${C_1}$\}? \\
  CQ & Identification & Which  \{${C\text{ name}}$\} has the highest/lowest \{${Q\text{ name}}$\}? \\
  CQ & Comparison & What is the difference of \{${Q\text{ name}}$\} between \{${C_1}$\} and  \{${C_2}$\}? \\
  CQ & Summation & What is the sum of \{${Q\text{ name}}$\} of \{${C_1}$\} and  \{${C_2}$\}? \\
  CTQ & Identification & What is the \{${Q\text{ name}}$\} of \{${C_1}$\} in  \{${T_1}$\}? \\
  CTQ & Identification & Which  \{${C\text{ name}}$\} has the highest/lowest \{${Q\text{ name}}$\} in  \{${T_1}$\}? \\
  CTQ & Identification & Which  \{${C\text{ name}}$\} has the highest/lowest \{${Q\text{ name}}$\} from  \{${T_1}$\} to  \{${T_2}$\}? \\
  CTQ & Comparison & What is the difference of \{${Q\text{ name}}$\} between \{${C_1}$\} and  \{${C_2}$\} in  \{${T_1}$\}? \\
  CTQ & Comparison & What is the difference of \{${Q\text{ name}}$\} between \{${C_1}$\} and  \{${C_2}$\} from  \{${T_1}$\} to  \{${T_2}$\}? \\
  CTQ & Trend Analysis & What is the trend of the \{${Q\text{ name}}$\} of \{${C_1}$\}? \\
  CTQ & Trend Analysis & What is the trend of the \{${Q\text{ name}}$\} of \{${C_1}$\} from  \{${T_1}$\} to  \{${T_2}$\}? \\
  CTQ & Summation & What is the sum of \{${Q\text{ name}}$\} of \{${C_1}$\} and  \{${C_2}$\} from  \{${T_1}$\} to  \{${T_2}$\}? \\
  CQQ & Summation & What is the sum of \{${Q\text{ name}}$\} of \{${C_1}$\} and  \{${C_2}$\} from  \{${T_1}$\} to  \{${T_2}$\}? \\
  \hline
  
  \end{tabular}
\end{table}

\subsection{Modeling Construction and Training} 

The task of this model is to identify the output attributes and the presence of comparison and derivation in natural language queries, which often have a nested structure that defies simple pattern matching.
To overcome this challenge, we leverage a large language model as the backbone and fine-tune it using a curated dataset of natural language to visualization task pairs.
This approach exploits the powerful transfer learning capabilities of large pre-trained language models.
In the following paragraphs, we describe the training requirements and our strategy for constructing the dataset.
\begin{itemize}[leftmargin=*]
  \item \textbf{R1.} Given NL inputs, the model should be able to predict the task type, for example, identity, comparison, and summarization.
  \item \textbf{R2.} Given natural language queries, the model should be able to predict whether there are filters on the attributes. If there is, the model should be able to determine the filtering operation (e.g., greater than, less than) and the filtering value (such as a categorical choice, temporal range, or value range).
  \item \textbf{R3.} Given natural language queries, the model should be able to determine whether there is a derivation on the attributes. If there is, the model should be able to predict the type of derivation.
\end{itemize}
We expect the model to have the following abilities: task operation prediction, referent extraction, derivation prediction, and attribute filter prediction.

\begin{figure*}[!ht]
  \centering
  \includegraphics[width=\textwidth]{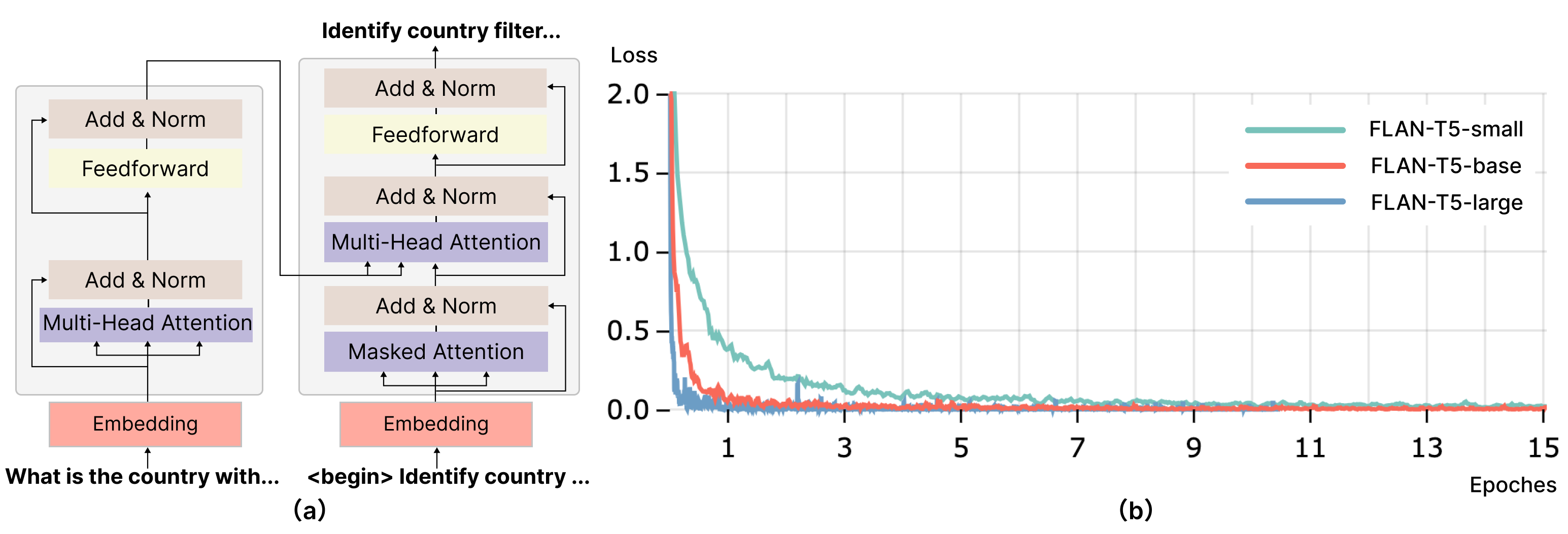}
  \caption{(a) The encoder-decoder model of the large language model, text-to-text translate transformer (T5~\cite{raffel2020exploring}).
  (b) The loss variation during the training process can be observed, showing that the model gradually converged over 30 epochs.}
  \label{fig:modelstructure}
\end{figure*}

\begin{itemize}[leftmargin=*]
\item \textbf{Task operation prediction} involves determining the operation type, for example, identifying and comparison, from the input natural language utterance.
\item \revision{\textbf{Referent extraction} is the capacity to extract focused visual elements from natural language, wherein these visual elements may be indicated through data attributes or visual channels. The model is required to discern and extract corresponding expressions from these referential cues.}
\item \textbf{Derivation prediction} involves predicting the query result based on the input natural language utterance and the attributes information.
For example, given the input ``What are the countries with the top-10 GDP per capita?'' and the attribute release year, the derivation ``rank'' should be extracted.
\item \textbf{Filtering prediction} aims to the filtering parameters, including the filtering operation, range, and filtering direction.
Filtering operations have four types: \texttt{equal to}, \texttt{larger}, \texttt{smaller}, and \texttt{inter range}.
Additionally, for ranking or position, there may be a filter direction, whether it is for the top, bottom, left, or right.
For example, ``What are the countries with the largest three populations'' has the filter ``operation: \texttt{<}, value: 3, direction: top.''
\end{itemize}
Translating natural language queries into nested hierarchical tasks can be regarded as a sequence-to-sequence translation process. Large language models, such as ChatGPT, have already demonstrated their capabilities in this kind of translation.
We use ChatGPT to construct a training dataset and leverage its ability to train a smaller LLM.
Taking into account the capabilities and size of large language models, we choose T5 as our training model.
One of the most powerful transformers, T5~\cite{raffel2020exploring}, which is short for \textbf{T}ext-\textbf{T}o-\textbf{T}ext \textbf{T}ransfer \textbf{T}ransformer, accepts sequence input and generates sequence output.
The T5 model is pre-trained on a large corpus of natural language. The model has an encoder-decoder structure, as shown in \autoref{fig:modelstructure} (a), where it takes input and generates output by predicting the probability of the next token given the input and the previous words in the output sequence. 
During fine-tuning, the T5 model masks the following words in the output sequence and predicts the output tokens.
Previous studies have demonstrated the effectiveness of fine-tuning the pre-trained T5 model in adapting to various downstream tasks~\cite{raffel2020exploring}.

Our method aims to endow the model with the generic ability to handle natural language queries for visualization.
However, collecting a large labeled training dataset is costly and challenging.
Therefore, we adopt a strategy of fine-tuning our tasks based on a pre-trained model.
This strategy enables the model to acquire new knowledge in the visualization tasks while preserving the learned natural language knowledge from the large-scale corpus. 
Based on the natural language transformer T5, we train the model to generate structured tasks from the natural language input.
An example of the input and output is as follows:
\begin{itemize}[leftmargin=*]
  \item \textbf{Input:}  \texttt{What is the consumption of coal in 2022?} 
  \item \textbf{Output:}
  \texttt{(identify consumption; filter: energy = coal, time = 2022)}.
\end{itemize}
We trained the model on the constructed dataset using an RTX Titan graphics card with 24 GB of memory.
We trained the model with different parameter sizes: small (60.5 million), base (223 million), and large (738 million).
Figure~\ref{fig:modelstructure} (b) presents the training loss for models with different parameters.

\section{Visualization Manipulation}
\label{section: manipulation}

Our approach primarily focuses on preserving user cognitive continuity with minimal mental shifts by predominantly utilizing manipulations based on existing visual elements. This approach offers two key advantages: (1) users can conserve more cognitive resources, and (2) our method can be applied to a wider array of visualizations. These manipulations, which modify the existing visual elements, align with Munzner's definition of manipulations~\cite{munzner2014visualization}.

\subsection{Design Space of Visualization Manipulation}

We categorize visualization manipulations based on the extent of changes made to the elements, taking into account factors including whether the elements' positions are changed, whether elements are added or removed, and whether the current encoding method is maintained.
We summarized the visualization manipulations into four levels and seven types, as depicted in \autoref{fig:visoperation}.
Here are the four levels of visualization manipulations:

\begin{figure*}[!htb]
    \centering
    \includegraphics[width=\textwidth]{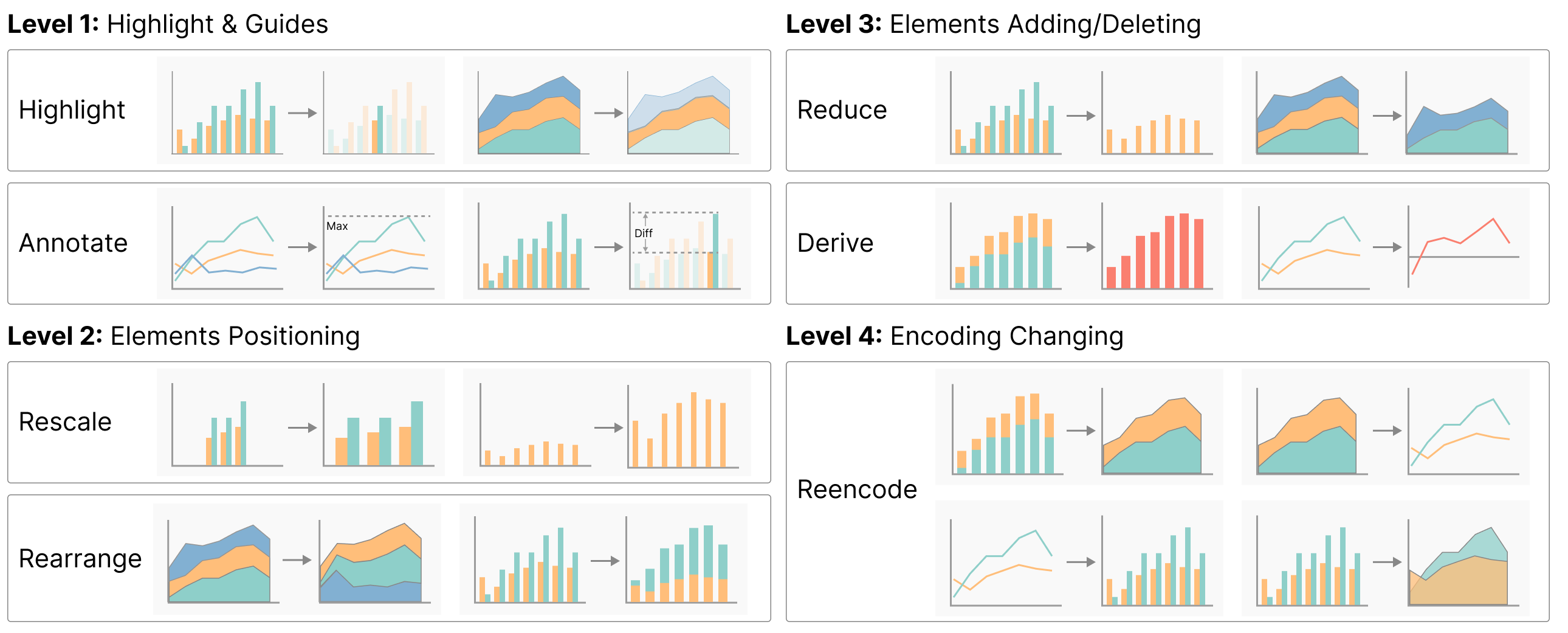}
    \caption{There are four levels and seven manipulations in the design space of visualization manipulation. 
    }
    \label{fig:visoperation}
\end{figure*}

\begin{itemize}[leftmargin=*]
    \item \textbf{Level 1. Highlight and Guides} modify highlights or introduce guidelines to make specific elements in the visualization prominent, directing the viewer's attention to key information.
    \item \textbf{Level 2. Element Positioning} adjusts the position of visual elements, changing their arrangement or layout to better facilitate the user's analytical tasks.
    \item \textbf{Level 3. Elements Adding/Deleting} add or remove visual elements in the visualization to emphasize or de-emphasize certain information or create derived visual elements.
    \item \textbf{Level 4. Encoding Changing} modifies the data encoding method (e.g., transforming a line chart into a bar chart) to accommodate user-specific queries.
\end{itemize}

The four levels of visualization manipulations encompass seven low-level types as depicted in \autoref{fig:visoperation}: 

\begin{itemize}[leftmargin=*]
    \item \textbf{Highlight} involves assigning varying visual intensities to existing elements, distinguishing between user-focused and non-focused content.
    \item \textbf{Annotate} adds auxiliary lines or text to the current visualization, emphasizing user-focused content. It is often used in conjunction with highlighting.
    \item \textbf{Rescale} adjusts the axis range to limit the range or achieve better space utilization after filtering, aligning, or stacking elements.
    \item \textbf{Rearrange} including alignment, stacking, and sorting, support comparison, summation, and ranking tasks, respectively.
    For instance, alignment manipulations force two or more visual elements to share the same baseline, typically for better comparison.
    Stack operations change (groups of) marks to display summarized results, transforming a grouped bar chart to a stacked bar chart or an overlapped area chart to a stacked one.
    Stack operations correspond to the summation of values for multiple data series.
    Sort operations reorder elements along the axis, facilitating the identification of rankings and supporting comparison.
    \item \textbf{Reduce} operation selectively retains focused visual elements while eliminating unfocused ones.
    \item \textbf{Derive} calculates new elements based on existing ones, placing them within the current visualization without changing the encoding. Examples include calculating the sum or difference between two elements.
    \item \textbf{Re-encode} manipulations modify the encoding of visual marks according to users' tasks, for instance, converting a line chart to an area chart or a bar chart.
    The re-encode space is vast, and we aim to minimize the understanding burden by avoiding encoding changes unless necessary.
    An example of a necessary change is when users require the summation results of a multi-line chart; re-encoding from a line chart to an area chart is essential, as stacking is not available for a multi-line chart.
\end{itemize}

Inspired by Liu et al.~\cite{liu2023spatialconstraint}, we introduced a design space of visualization manipulation that employs control points and spatial constraints to represent visual elements.
Compared to Liu et al.~\cite{liu2023spatialconstraint}, which can only support manipulations that do not alter the number of control points, number of visual elements, and visualization type, our approach allows for annotation, derivation, and reencoding.
Our aim is to augment the users' comprehension of their tasks.
We refrain from implementing modifications that would fundamentally transform the visualization.
Modifications to the encoding can be achieved through the addition or removal of control points and the appropriate adjustment of constraints.
It is essential that all modifications are based on the information already present within the original visualization and support modifications starting from the original visualization.

\subsection{Mapping Tasks to Manipulations}

This section discusses how visualization tasks are transformed into visualization manipulations to respond to user queries.
Visualization manipulations serve as a response to visualization tasks in user queries, providing a smooth transition from the current visualization layout to a new one.
We explain how different tasks are converted into visualization manipulations based on the hierarchical structure-tasks summarized in \autoref{section: nl}.

\begin{itemize}[leftmargin=*]
    \item \textbf{Filter} tasks are fulfilled by selecting focused visual elements to display based on specific attribute values. We process filter tasks in a bottom-up manner. We handle filter tasks with \textbf{highlight} and \textbf{reduce} manipulations, depending on the number of filtered elements.
    When the number of selected data items is small, we present the context information by highlighting the selected ones while keeping the other filters visible. After completing the \textbf{reduce} operation, the range of data for visual elements in the chart may change, which can reduce screen utilization. To address this, we perform a \textbf{rescale} operation to adjust the display range accordingly.
    
    \item \textbf{Derivation} tasks involve generating new attributes based on the original attributes.
    We process derivation tasks after filter tasks. 
    When performing differences or sums, we use the \textbf{derive} operation to add new visual elements resulting from calculations involving two or more visual elements. Users can then ask questions about these new elements.
    
    \item \textbf{Identification} tasks consist of a certain number of filter and derivation tasks. After filtering and deriving results, the corresponding results are highlighted and annotated within the visualization to address the user's question.

    \item \textbf{Comparison} tasks involve combinations of several identification tasks. Comparison tasks are carried out by simultaneously displaying multiple identification tasks, enabling users to observe contrasting effects.

    \item \textbf{Aggregation} tasks involve generating a single value from a value list given an operation, such as extremes and averages.
    \item \textbf{Annotation} manipulations are applied to the visualization to assist users in identifying specific numerical characteristics.

\end{itemize}

In summary, when transforming tasks into visualization manipulations, we adhere to several principles: executing visualization manipulations in a \textbf{bottom-up order}, displaying necessary context, and minimizing re-encoding manipulations. We first perform filtering based on the tasks' included filters, deciding whether to highlight or remove elements depending on the number of visual elements remaining. If there are too few remaining visual elements, we display context information (e.g., information from adjacent time periods). 
Then, we strive to minimize re-encoding manipulations and maintain consistency in visual form. When the current visualization is insufficient to address a task, we perform a re-encode operation to better answer the user's question.

\begin{figure}[!ht]
    \centering
    \includegraphics[width=\columnwidth]{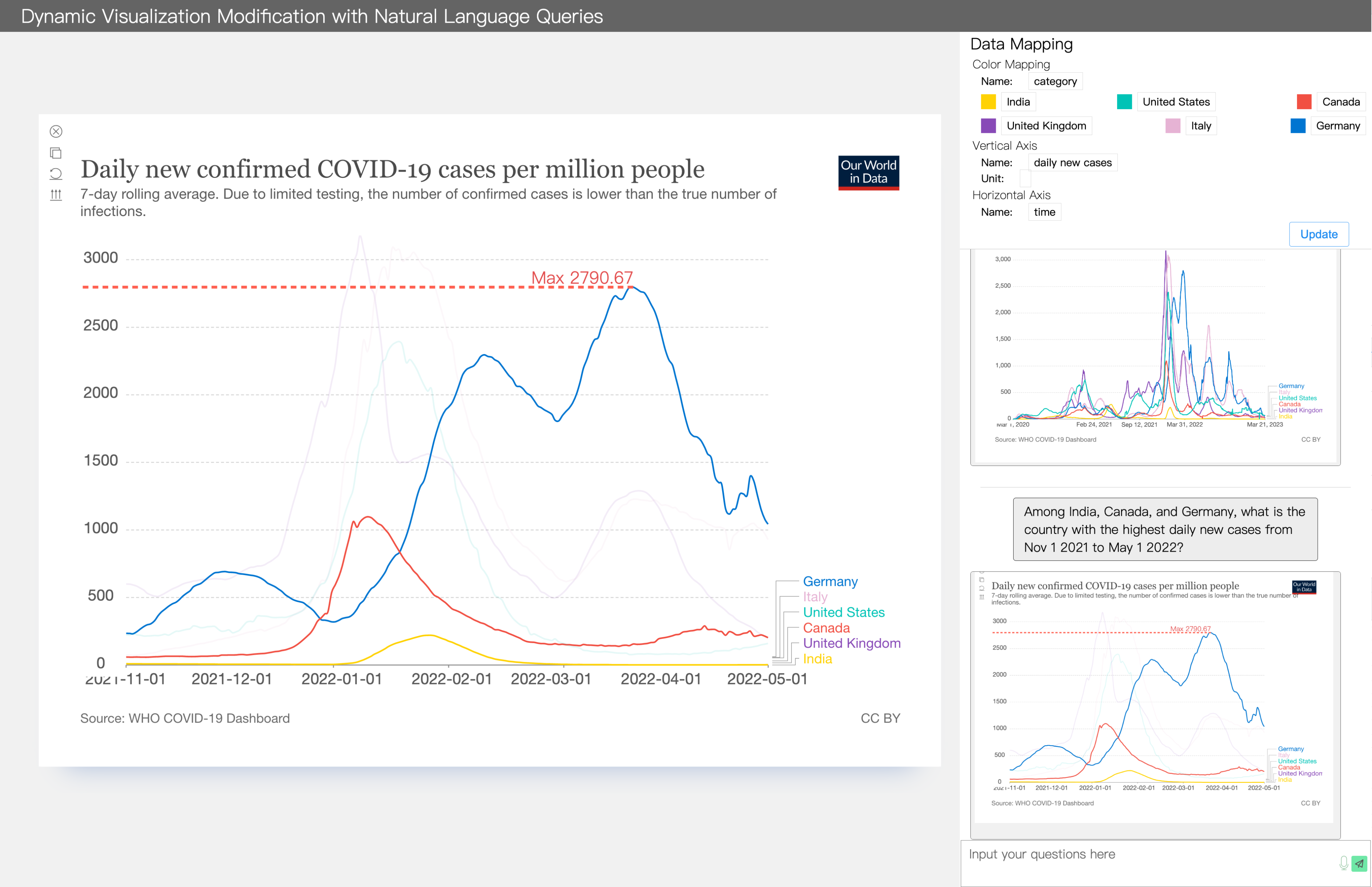}
    \caption{The interface of our system. Users can upload a visualization chart, and use natural language queries to interact with the system.}
    \label{fig:interface}
    \vspace{-10px}
\end{figure}

\begin{figure*}[!ht]
    \centering
    \includegraphics[width=\textwidth]{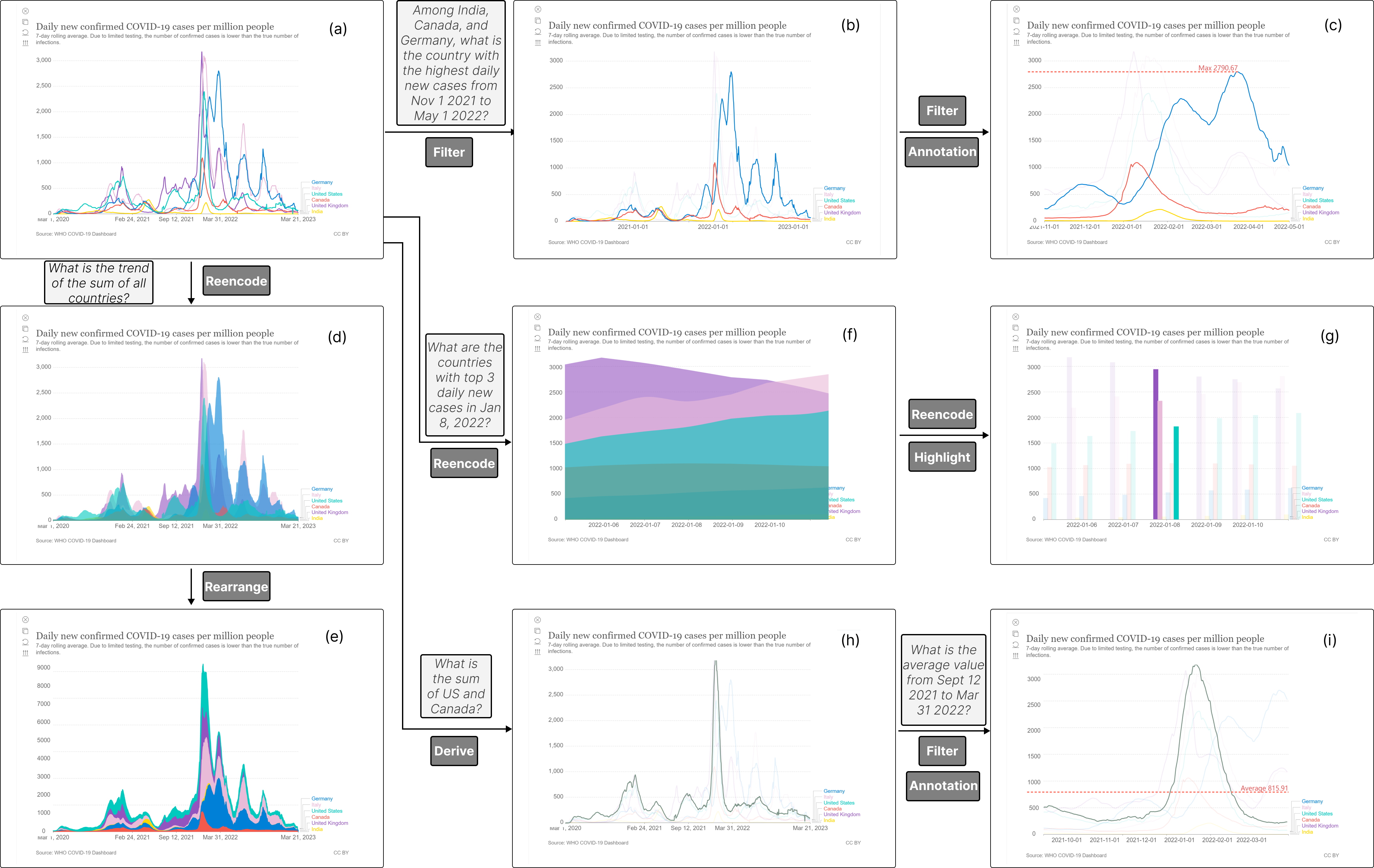}
    \caption{After parsing user-uploaded visualizations, various natural language questions can be posed by the user. These queries may involve different filtering conditions on multiple attributes, as well as derivative content. The corresponding visualizations will undergo multiple transformations to answer the user's questions. Throughout this process, our system aims to maintain continuity between visualizations, and in cases where the current visualization is insufficient to support the users' query, our system transitions to other types of visualizations to support the user's query.}
    \label{fig:case}
\end{figure*}

\section{Usage Scenario} 
\label{section:use_scenario}

\autoref{fig:case} displays a visualization from Our World in Data website\footnote{Our world in data: \url{https://ourworldindata.org/}}. This visualization illustrates the daily increase of COVID-19 cases in six different countries over a three-year period, consisting of more than 6,000 data points. We chose this visualization as our case due to its diverse attributes, including time, categorical, and quantitative properties. Users exploring this visualization face challenges such as overly dense lines and compact data points, making it difficult to examine specific time steps.

Taking an SVG-format visualization as input, our system enables seamless and continuous exploration.
Users can pose various queries atop the system, as shown in \autoref{fig:interface}, and the system naturally transforms to support different queries.
Starting from \autoref{fig:case} (a), users can ask a variety of tasks, and the visualization adapts in response to the users' natural language queries. After users input their queries in natural language, our system manipulates the visualization through animations to answer the question, showcasing both intermediate steps and final results. As depicted in \autoref{fig:case} (b) and (c), when the user asks, ``Among India, Canada, and Germany, what is the country with the highest daily new cases from Nov 1, 2021 to May 1, 2022?'', we first extract the filters and apply highlighting or filtering manipulations. In this case, we use highlighting for the categorical attributes and filtering and rescaling for the time filter. Finally, we use an annotation manipulation for the max value task in the derived tasks, arriving at the final result.

As shown in \autoref{fig:case} (d) and (e), when the user asks about the overall trend for all countries, area charts hold a significant advantage over line charts in presenting the sum trend.
Therefore, we first employ the reencode manipulation to convert the line chart to an area chart, then use the rearrange manipulation to transform it into a stacked area chart, displaying the combined trend for all countries. As illustrated in \autoref{fig:case} (f) and (g), when the user inquires about the rankings at a specific point in time, bar charts better represent the data for a single date. Accordingly, we first apply a filter and rescale manipulation to focus on the time frame and convert the visualization to a bar chart using the reencode manipulation. Subsequently, we calculate the rank value derived from the manipulation and apply a highlight manipulation based on it to obtain the final chart that answers the question.

In \autoref{fig:case} (h) and (i), the user initially wants to view the combined trend of the two countries.
We employ a deriving manipulation to obtain a line representing the sum of the two countries and highlight it.
Following this inquiry, the user desires the average value of the sum of the two countries within a specific time frame.
We first showcase the time range using filter and rescale manipulations, then calculate the average value and add an annotation.

These examples demonstrate our system's support for various types of user questions. By parsing questions into tasks using the model, our system employs distinct visualization manipulations to create continuous animated transitions from the original chart, ultimately generating the final chart to answer users' queries and meet their needs.

\section{Evaluation}
\label{section:userstudy}

This section includes a quantitative evaluation and a user study.

\subsection{Quantitative Evaluation}

In our quantitative assessment of the model's accuracy in parsing natural language tasks, we employ five distinct metrics, namely literal, semantic, task, filter, and format accuracy.
\begin{itemize}[leftmargin=*]
\item \textbf{Literal accuracy} refers to whether the output matches the ground truth in the string level.
Difference in the order of keys and list items in the structure is tolerated.
The elements of the actual structure and predicted structure are sorted alphabetically and then compared to calculate the literal accuracy.

\item \textbf{Semantic accuracy} refers to the equivalence between the predicted structure and the actual one in terms of denoting tasks.
Specifically, in some sentences, multiple attributes are mentioned through visual channels with brief expressions like ``the green and blue lines.''
In such cases, whether the filter corresponding to the green object includes ``shape'' or not does not affect the extraction of the object and can be considered semantically correct.

\item \textbf{Task accuracy} refers to whether the task is correctly predicted.

\item \textbf{Filter accuracy} refers to the ratio of the correctly predicted filters.
Sometimes, the model might accurately predict certain filters while making incorrect predictions for others.
We use a value between 0 and 1 to represent the proportion of filters that are correctly predicted.

\item \textbf{Format accuracy} refers to whether the predicted value is in the correct format and can be correctly parsed to the structure.
The most common incorrect format involves missing brackets.
\end{itemize}

\begin{table*}[htbp]
    \renewcommand{\arraystretch}{1.2}
    \newcolumntype{Y}{>{\centering\arraybackslash}X}
    \newcolumntype{C}{>{\arraybackslash}p{3.5cm}}
    \centering
    \caption{Qualitative Results}
    \label{tab:benchmark}
    \begin{tabularx}{0.95\textwidth}{C@{\hspace{0.3cm}}*{5}{Y}}
      \hline
      \multicolumn{1}{c}{\textbf{Model}} & \textbf{Literal (\%)} & \textbf{Semantic (\%)} & \textbf{Task (\%)} & \textbf{Filter (\%)} & \textbf{Format (\%)}  \\
      \hline
      $\text{FLAN-T5-small}$ \hfill $\text{ }_{\text{(5\  epochs)}}$ & 69.792 & 73.611 & 99.306 & 88.468 & 99.653 \\
      $\text{FLAN-T5-small}$ \hfill $\text{ }_{\text{(10 epochs)}}$ & 86.111 & 89.583 & 98.958 & 95.045 & 99.306 \\
      $\text{FLAN-T5-small}$ \hfill $\text{ }_{\text{(15 epochs)}}$ & 88.889 & 91.667 & 99.306 & 96.421 & 99.653 \\
      $\text{FLAN-T5-base\ \ }$ \hfill $\text{ }_{\text{(5\  epochs)}}$ & 87.847 & \textbf{92.708} & \textbf{99.653} & 96.668 & \textbf{100.0} \\
      $\text{FLAN-T5-base\ \ }$ \hfill $\text{ }_{\text{(10 epochs)}}$ & 84.028 & 88.194 & 98.958 & \textbf{97.558} & 99.306 \\
      $\text{FLAN-T5-base\ \ }$ \hfill $\text{ }_{\text{(15 epochs)}}$ & 87.153 & 90.972 & 97.917 & 96.675 & 98.264 \\
      $\text{FLAN-T5-large}$ \hfill $\text{ }_{\text{(5\  epochs)}}$ & 85.764 & 90.625 & 97.569 & 95.844 & 97.917 \\
      $\text{FLAN-T5-large}$ \hfill $\text{ }_{\text{(10 epochs)}}$ & 85.764 & 89.236 & 98.958 & 96.458 & 99.306 \\
      $\text{FLAN-T5-large}$ \hfill $\text{ }_{\text{(15 epochs)}}$ & \textbf{90.625} & 92.014 & 98.611 & 95.602 & 98.958 \\
        \hline
    \end{tabularx}
  \end{table*}

We compare the parsing accuracy of large language models with varying parameter sizes across three different epochs (5, 10, 15), as presented in \autoref{tab:benchmark}. It is evident that models with different parameter sizes perform proficiently in extracting tasks from natural language.
Notably, the model with the base size demonstrates excellent capability in handling the tasks described in this study.

\subsection{User Study}

\textbf{Participants.}
We recruited 10 participants, which are undergraduate or graduate students from various majors, including 3 females.
Participants are required to rate their experience with charting software (e.g., Excel, Tableau) and programming tools (e.g., $D^3$, Vega) on a 5-point Likert scale, where 1 denotes ``never heard of'' and 5 denotes ``very familiar.''
Their responses indicated a diverse range of experience in data visualization, i.e., charting software ($\mu=3.9$, $\sigma=1.10$) and programming tools ($\mu=4.6$, $\sigma=0.70$).

\textbf{Procedure.} We utilize the visualization from Our World in Data as introduced in \autoref{section:use_scenario}.
We allow users to explore and interact with the visualizations through natural language queries that the visualizations could answer. After familiarizing themselves with the system, the users interacted with the visualizations for approximately 30 minutes.

\textbf{Interview Questions.} We inquired about the advantages and disadvantages of each feature and the participants' overall opinions of the system. Then, we asked them to rate the system in terms of natural language parsing accuracy (NL Parsing Accuracy), the reasonableness of visualization manipulation changes (VIS Operation Rationale), the degree of support for exploration (Exploration Support), and overall utility (Overall Utility).

\begin{figure}[!ht]
    \centering
    \includegraphics[width=\columnwidth]{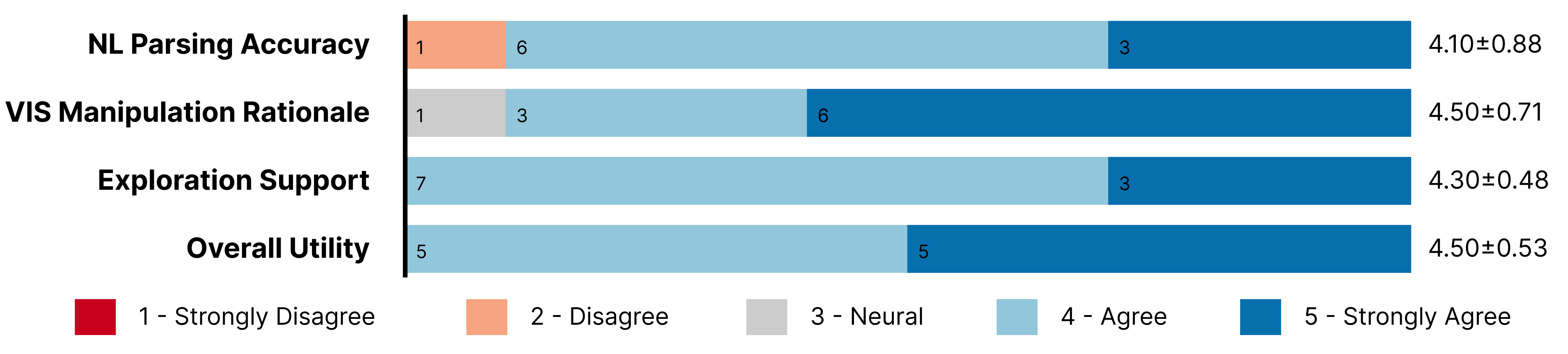}
    \caption{Users' ratings for our system include the accuracy of natural language parsing, the rationality of visualization manipulations, support for the exploration process, and overall rating.}
    \label{fig: userstudy}
\end{figure}

\textbf{User Feedback.}
Overall, our system has received relatively high evaluations from participants.
As shown in Figure \ref{fig: userstudy}, participants generally found the system to be quite useful ($\mu = 4.5$, $\sigma=0.53$). In the following text, we summarize the feedback from participants regarding the strengths of the system.

\begin{itemize}[leftmargin=*]
   \item  \textit{Accurate natural language parsing:} 
   Most users consider the natural language parsing to be accurate ($\mu = 4.1$, $\sigma=0.88$).
   The system is able to accurately recognize the input natural language and convert the aspects of interest into visual charts (P2). Some users emphasized that the system supports fuzzy queries or the use of simple words as filters (P8), which helps broaden the scope of user interaction. The natural language parsing accuracy of the system is considered high (P4, P5).

    \item \textit{Intuitive animation and transition:}
    The visualization manipulations are regarded as rational ($\mu = 4.5$, $\sigma=0.71$).
    Smooth animations display the process of visualization changes, aiding in the understanding of the calculation process and data connection (P3, P6, P8, P10). The system also supports the transformation of various visualization forms, allowing users to understand data from multiple perspectives (P5, P9).

    \item \textit{Flexibility and continuous exploration:} 
    The system is considered to be useful to support the exploration ($\mu = 4.3$, $\sigma=0.48$).
    The system supports the parsing of nested natural language queries, enabling continuous changes across multiple questions (P1). Moreover, the system facilitates the exploration of more complex questions through multi-step transformations (P7).
    P2 stressed that \textit{``Converting aspects of user interest into visual representations, which can then be transformed into various types of visual charts, allows users to gain multifaceted insights into the data presented in the charts.''}

    \item \textit{Interactivity and user-friendliness:} 
    Based on natural language, users can interact with the system in real-time, helping to identify chart content and highlight areas of focus (P4).
    Users can also directly control the visualization through natural language, providing accurate feedback for clear instructions (P10). The system offers an aesthetically pleasing interface and aligns with users' intuitiveness.
    P10 also appreciated the design of the chat box, which made him feel like chatting with an agent. He can revisit chat and result history easily, helping him gather all information to establish a wholesome understanding of the visualization.

    \item \textit{Direct manipulation:}
    During the user study, a participant (P4) asks the system to zoom in to take a closer look. However, the system can not understand the zoom instruction of the vagueness to interpret scale and focus parameters.
    Finally, P4 solved the problem by filtering the focal area under our guidance and complained that the filter operation was not as intuitive as the zoom operation. 
    The limitation arises from the fact that natural language is not always the most convenient option in every scenario, as some user tasks may lack clear directives, such as when users are freely exploring.
    The combination of direct manipulation and natural language manipulation might enable more flexibility.

    \item \textit{Open world knowledge:}
    Two participants (P5, P7) posed tasks that required open-world knowledge to complete, such as \textit{``What is the sum of confirmed cases in countries across different continents.''} The system failed to group the countries by inferring the relationship between countries and continents.
    The size of our current model limits its ability to harness open-world knowledge acquired during pre-training.
    However, this limitation can be overcome by utilizing a larger language model or incorporating external knowledge databases.

    \item \textit{Deconstructing high-level expressions}: 
    Another type of failure case requested by users involves high-level tasks. For example, P6 asked the question, \textit{``Which period do all countries suffer from the most concentrated pandemic outbreak?''} This request failed due to the system's limitation in deconstructing the high-level expression ``most concentrated outbreak'' into ``finding maxima'' for ``outbreak'' and ``comparing closeness'' for ``most concentrated.'' The enhanced inference capabilities of large language models might facilitate the interpretation and deconstruction of expressions with high-level semantics.

\end{itemize}

\section{Discussion and Future Work}
\label{section:discussions}

In this section, we discuss the limitations of our model and future work that we plan to pursue. While our current model enables user-driven visualization to answer questions, there is still room for improvement in the following areas:

\subsection{Expanding the Range of Visualizations}
At present, our method relies on reverse engineering to correct or provide underlying data, and we primarily focus on bar charts, line charts, and area charts, which are the most common types of visualizations~\cite{battle2018beagle}.

In the future, we aim to support a wider range of visualizations. Extracting accurate information from complex visualizations is a challenging task that may require a combination of interactive and intelligent approaches. Therefore, we will develop a more robust reverse-engineering method that integrates interaction and intelligent methods to obtain underlying data from visualizations.

\subsection{Towards Multi-Modal Interaction}

A few participants have expressed that natural language may not always be the most efficient way to convey certain queries, especially those related to precise timing. Descriptions in natural language can be lengthy and may face issues with machine parsing inaccuracies. Additionally, some interactions cannot be easily represented using the traditional Window, Icon, Menu, and Pointer (WIMP) interface.
To address these limitations, we plan to combine the strengths of both approaches and develop efficient multi-modal interaction systems. Natural language can also be used as a programming language for traditional interfaces, allowing complex interactions to be mapped to natural language using simple languages.

\subsection{Customization for Different Users}

Currently, our model operates on a set of pre-established rules for visualizing natural language tasks. However, However, different users may have their own preferences. For instance, when querying the sum of multiple lines on a multi-line chart, some users may prefer to see the summed result represented as a single line, while others may prefer stacked areas. These preferences arise due to varying expectations regarding chart types and the level of detail provided by visualizations. To address this, we plan to incorporate user feedback systems to allow for the integration of user preferences into the visualization generation process.

\subsection{NLI for Complex Visual Analytic Systems}

Our current work focuses on visualizations based on common chart types. Developing a natural language interface for more complex, interactive visual analytic systems is challenging due to the diverse tasks and varying semantics associated with different data and tasks. In the future, we can explore the design space of visual analytic systems and extract frequently used natural language patterns and custom modules. Additionally, we aim to leverage large-scale language models to rapidly build natural language interface plug-ins for existing visual analytic systems.

For a broader range of visualizations, creators can use natural language to specify the semantics of visualization components and features, which can significantly enhance usability.
For instance, creators can define features to represent specific areas of focus within a domain.
Once published, users can engage in highly customized natural language interactions.
We aim to provide a universal natural language interaction framework that supports custom modules for various visual analytic systems.
This framework can be fine-tuned through few-shot or zero-shot learning on these custom modules, enabling seamless integration and adaptation to different contexts and visualization requirements.

\section{Conclusion}
\label{section:conclusions}

In this paper, we propose a pipeline for seamlessly manipulating existing visualizations to answer users' natural language queries. 
To achieve this, we first analyze the design space of visualization tasks. 
We fine-tune a large language model to extract hierarchical tasks inherent in these natural language queries.
To train the model, we curated a dataset with cross-domain data attributes, various tasks, and multifaceted expressions with the help of LLM.
Using this dataset, we train a natural language-to-task translator that can extract hierarchical tasks from various natural language queries.
These tasks are subsequently utilized to execute a series of manipulation operations on the visualizations,
We evaluate our method quantitatively and qualitatively and demonstrate that the natural language-to-visualization task translator accurately extracts task information and the manipulated results effectively help users understand the tasks.

\bibliographystyle{ACM-Reference-Format}
\bibliography{manuscript}

\end{document}